\documentclass{caosp305}
\usepackage{graphicx}
\usepackage{natbib}
\usepackage{ulem}

\articleNo{IBWS01}
\pubyear{2017}
\volume{47}
\volnumber{2}
\firstpage{1}
\received{May 5, 2017}
\accepted{May 22, 2017}   

\def\BibTeX{{\rm B\kern-.05em{\sc i\kern-.025em b}\kern-.08em
             T\kern-.1667em\lower.7ex\hbox{E}\kern-.125emX}}

\begin{document}

\hauthor{Vladim\'{\i}r Karas et al.}

\title{Plunging neutron stars as origin of organised magnetic field in galactic nuclei}

\author{
        V.\,Karas \inst{1} 
      \and
        O. Kop\'a\v{c}ek \inst{1}  
      \and
        D. Kunneriath \inst{2}
      \and
        M.\,Zaja\v{c}ek \inst{3,} \inst{4}      
      \and \\[3pt]
        A.\,Araudo \inst{1}
      \and
        A.\,Eckart \inst{3,} \inst{4}
      \and 
      J.\,Kov\'a\v{r} \inst{5}
   }

\institute{Astronomical Institute of the Czech Academy of Sciences,\\
 Bo\v{c}n\'{\i} II 1401, CZ-14100 Prague, Czech Republic\\ \email{vladimir.karas@cuni.cz}
         \and
 National Radio Astronomy Observatory,\\ 520 Edgemont Road, Charlottesville, VA 22903, USA
         \and
 I. Physikalisches Institut der Universit\"at zu K\"oln,\\ 
   Z\"ulpicher Strasse 77, D-50937 Cologne, Germany
         \and
 Max-Planck-Institut f\"ur Radioastronomie,\\ Auf dem H\"ugel 69, D-53121 Bonn, Germany
         \and
 Faculty of Philosophy and Science, Silesian University in Opava,\\
        Bezru\v{c}ovo n\'am. 13, CZ-74601 Opava, Czech Republic
          }

\date{May 22, 2017}

\maketitle

\begin{abstract}
Black holes cannot support their own internal magnetic field like, for example, compact stars can. Despite this fact observations indicate that event horizons of supermassive black holes (SMBH) are threaded by field lines along which plasma streams flow. Various magnetohydrodynamical mechanisms have been suggested to generate turbulent magnetic fields on small scales, however, the origin of the large-scale component is unclear. In this write-up we describe our progress in an on-going work and discuss the possibility of dipole-type magnetic fields being brought onto SMBH by magnetized neutron stars, which are expected to drift inward from a hidden population in the Nuclear Star Cluster. This can contribute to an organised component of the magnetic field on the characteristic length-scale of the stellar size, which thread the horizon during the final stages of the magnetized star plunge into or its close flyby around SMBH.
Because of mass--size scaling relations for black holes, the effect is more important for lower-mass SMBH.
\end{abstract}

\section{Introduction}
Effects of mutual interaction between strong gravitational and electromagnetic fields play an important role in shaping
structures and driving processes in the inner regions of galactic nuclei. These are governed by the system of 
Einstein-Maxwell equations, which represent a highly non-linear set of coupled partial differential equations.
Special solutions can be derived by imposing various idealized assumptions and symmetries, however,
astrophysically realistic solutions can only be found by numerical approaches or approximative
methods, such as linearization.

In galactic nuclei, gravity is determined by the central black hole which dominates within its sphere of influence, $r\la r_{\rm s}$ \citep{Merritt2013},
\begin{equation}
 r_{\rm s}\simeq1.7\,\frac{M_\bullet}{4\times10^6M_{\odot}}\left(\frac{\sigma}{100\,{\rm km/s}}\right)^{-2}\,{\rm pc}, 
 \label{eq_sphere_influence}
\end{equation}
where $M_\bullet$ is the mass of the black hole expressed in units of the inferred mass of the Galactic centre black hole \citep[Sgr~A*; see][]{2010RvMP...82.3121G,Eckart2017} and $\sigma$ denotes the line-of-sight stellar velocity dispersion in the central region.

Further out, contributions to the gravitational field arise from stars in the Nuclear Star Cluster (NSC), as well as gas and dust in an accretion disc and a circum-nuclear torus. Nonetheless, the latter terms can be neglected close to the black hole horizon, where the solution of Einstein's field equations (in geometric units, $c=G=1$), is given by Kerr metric to high degree of accuracy,
\begin{equation}
R_{\alpha\beta}-\textstyle{\frac{1}{2}}Rg_{\alpha\beta}=8\pi T_{\alpha\beta}
\label{eq1}
\end{equation}
\citep[see][]{1973grav.book.....M}; here we assume zero electric charge of the black hole for simplicity, $Q=0$. 

Vacuum space-times have the right-hand side of eq.\ (\ref{eq1}) vanishing, whereas for a more general electro-vacuum case the energy-momentum tensor $T_{\alpha\beta}$ is determined by electromagnetic terms:
\begin{equation}
T^{\alpha\beta}\equiv T^{\alpha\beta}_{\rm EMG}=\frac{1}{4\pi}\left(F^{\alpha\mu}F^\beta_\mu- \textstyle{\frac{1}{4}}F^{\mu\nu}F_{\mu\nu}g^{\alpha\beta}\right) \simeq \mathcal{O}(\mathcal{E}^2+\mathcal{B}^2).
\label{eq2}
\end{equation}
The electromagnetic field tensor $F^{\alpha\beta}$ contributes to the source, however, astrophysically realistic fields are not strong enough and their gravitational influence to the space-time metric terms vanishes when linearized in electric ($\mathcal E$) and magnetic ($\mathcal B$) intensities \citep{1978JETP...47..419G,2010PhRvD..82h4034F}.

Gravitational radius, $r=r_{\rm g}=1.475\times10^5M/M_{\odot}\,{\rm cm}$, characterizes the typical length-scale in a system which involves action of strong gravity near a black hole. In the units of gravitational radius, the horizon of a rotating black holes has radius equal to $r_+(a)=1+\sqrt{1-a^2}$, which gradually decreases with black hole dimensionless spin $a$: $r_+=2$ for a non-rotating (Schwarzschild) black hole, and $r_+=1$ for a maximally spinning (Kerr) black hole. These represent asymptotically flat vacuum solutions endowed with a regular event horizon, which is present in the Kerr metric for $a^2\leq1$.

\begin{figure}[tbh!]
\centering
\includegraphics[width=\linewidth]{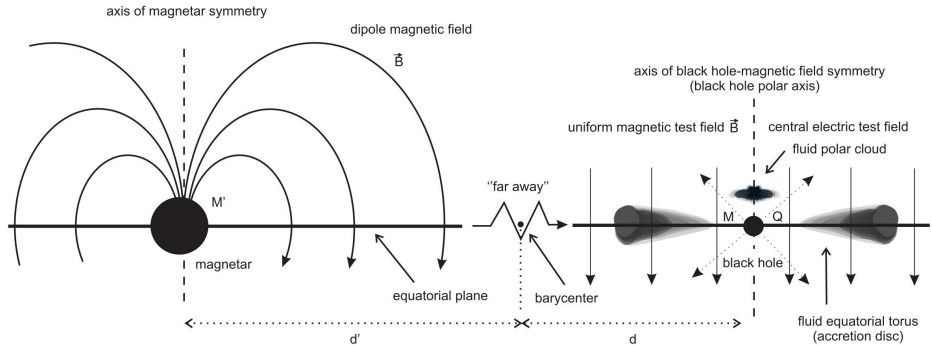}
\caption{An artistic sketch of a binary system of a magnetized neutron star interacting with a supermassive black hole. Magnetic field has dipolar topology near the surface of the neutron star (on the left side of the diagram), which we assumed to be non-rotating for simplicity. At larger distances, within a limited volume surrounding the companion black hole, the imposed magnetic field resembles a homogeneous filament of (almost) parallel lines \citep[on the right side; see][]{2014PhRvD..90d4029K}.}
\label{fig1}
\end{figure}

Let us note that special solutions to eqs. (\ref{eq1})--(\ref{eq2}) for electromagnetic test fields can be generated by 
symmetries of the underlying spacetime \citep{1974PhRvD..10.1680W}. In case of Kerr metric, the presence of two Killing vectors corresponds to the to stationarity and axial symmetry. Killing vectors satisfy the well-known equation \citep{1973grav.book.....M},
\begin{equation}
\xi_{\mu;\nu}+\xi_{\nu;\mu}=0,
\end{equation}
where coordinate system is selected in such a way that the following
condition is satisfied: 
$\xi^\mu=\delta^\mu_\rho$. 
Naturally, the problem is greatly simplified by assuming axial symmetry and stationarity, however, in full generality the electromagnetic field may or may not conform to the same symmetries as the gravitational field. In an asymptotically flat spacetime, the Killing vector $\xi\rightarrow\partial_\phi$ generates a uniform test magnetic field that is aligned with the rotation axis and perfectly homogeneous at large distance, whereas the field vanishes asymptotically for $\xi\rightarrow\partial_t$. It should be evident that the induced electric field vanishes in the non-rotating case. Furthermore, we remind the reader that material objects, namely, a self-gravitating torus and dusty clouds likely surround the black hole and they contribute to the gravitational field of any astrophysically realistic SMBH object \citep[e.g.][and references cited therein]{1987Natur.329..810S,2015ApJS..221...25P,2016ApJ...832...15C}. Therefore, self-consistent solutions must be more complicated.

Magnetic fields near accreting black holes operate on vastly different length-scales, $\ell$, which range from small-scale turbulent fields ($\ell\ll r_+$, likely due to magneto-rotational instability), to organized fields operating on larger scales exceeding the radius of horizon, $\ell\ga r_+$. In the present contribution, we are interested in the latter possibility. While it has been suggested that organized field lines could support the process of launching and pre-collimating jets, the origin of filamentary structures remains unclear. Different options have been discussed in the literature; here we propose that the horizon of a supermassive black hole may be threaded by a dipole-type magnetic component that originates in a strongly magnetized neutron star in its vicinity. A significant number of remnant (isolated) neutron stars is expected to exist in the Nuclear Star Cluster, where they can gradually sink towards the black hole.\footnote{Let us note that the presence of a magnetar has been reported in the neighbourhood of the Galactic centre \citep[SGR J1745-29; see][]{2013ApJ...770L..24K,2013ApJ...770L..23M,2013ApJ...775L..34R}, however, the distance $\simeq0.5\,$pc is too large to provide a convincing example of the system envisaged in our present discussion.}

Figure \ref{fig1} illustrates the basic idea about magnetic stars as the origin of external magnetic field. A similar set-up was proposed as a model for the origin of accretion disc coronae levitating because of the presence of an organized magnetic field above the equatorial plane and in polar regions near an accreting black hole \citep{2010CQGra..27m5006K,2014PhRvD..90d4029K}. Here we concentrate on the innermost parts of the system where the magnetic field threads the event horizon \citep[see also][]{2016EPJC...76...32S}.

\begin{figure}[tbh!]
\centering
\includegraphics[scale=0.345, clip,trim=30mm 5mm 67mm 5mm]{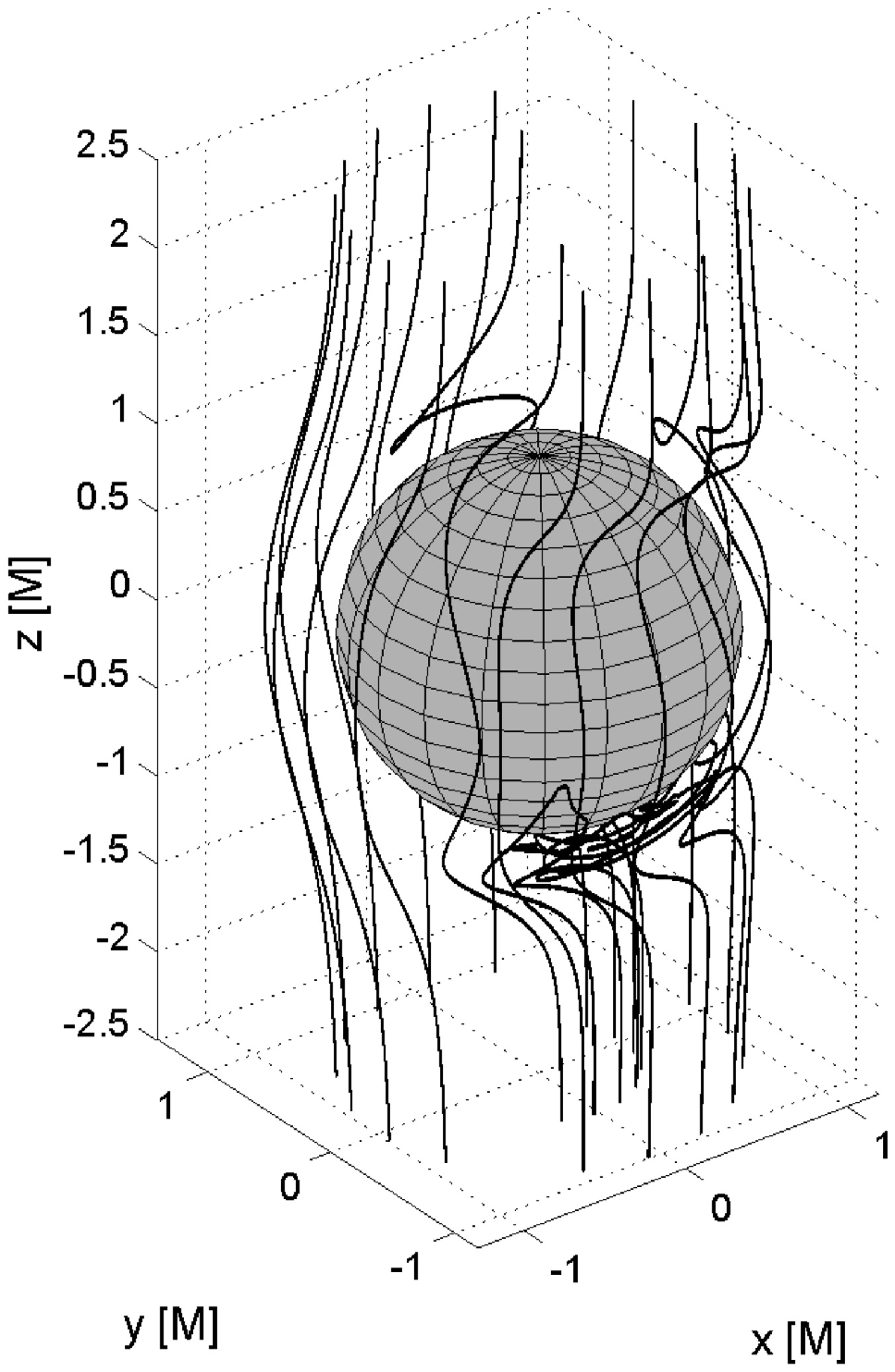}
\includegraphics[scale=0.345, clip,trim=56mm 5mm 65mm 5mm]{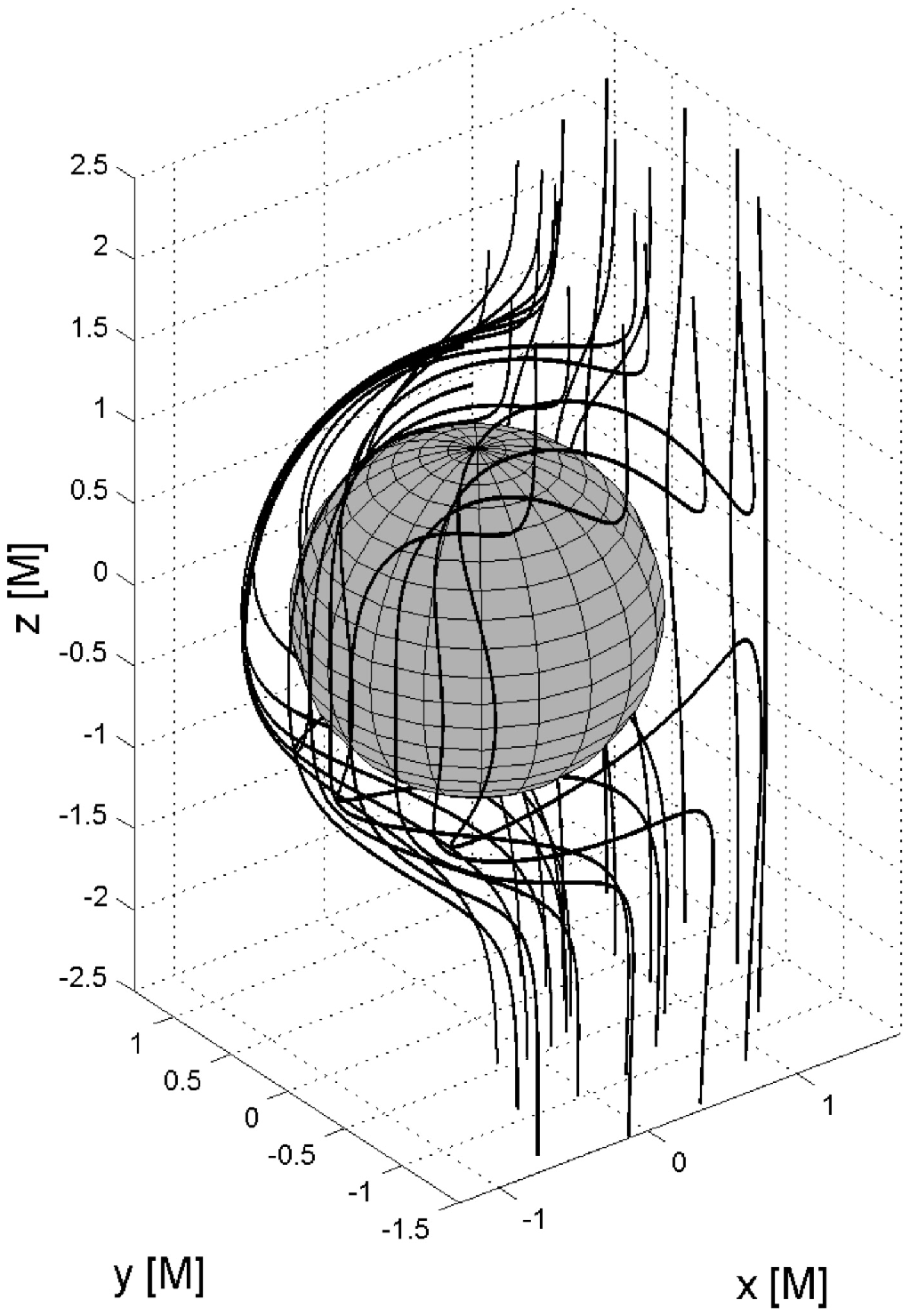}
\includegraphics[scale=0.345, clip,trim=55mm 5mm 65mm 5mm]{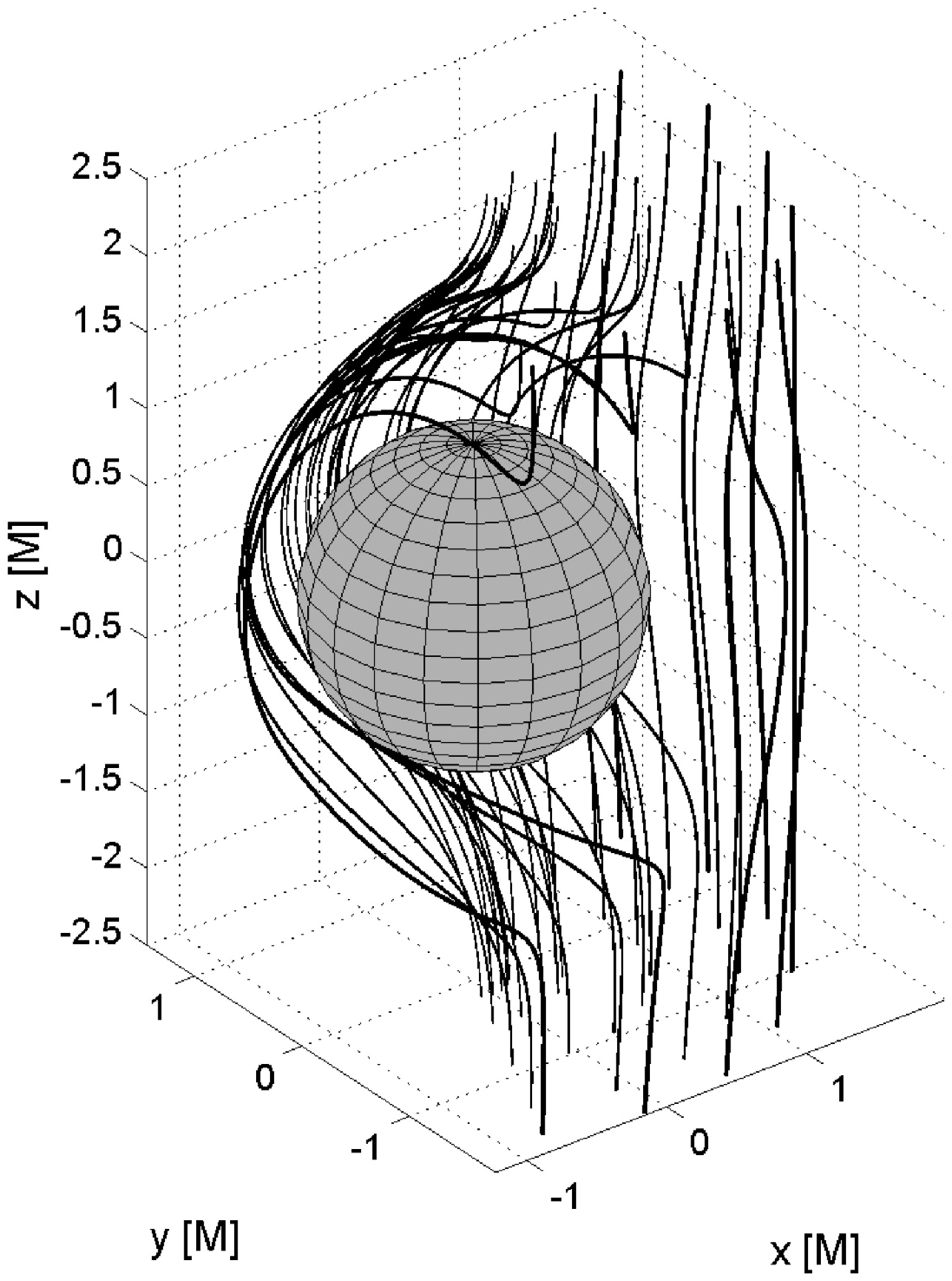}
\caption{Structure of asymptotically homogeneous (vertically aligned) magnetic field lines influenced by translatory boost of the black hole (horizon $r=r_+$ is indicated by the shaded sphere). The case of extreme rotation ($a=1$) of Kerr black hole is shown and the field lines are defined with respect to ZAMO observers. The field becomes expelled out of the horizon  (Meissner-type effect) for the vanishing drift velocity, but it can thread the horizon with an increasing flux when the black hole moves. The direction of boost velocity has been set to coincide with the horizontal axis of the stereometric projection: $v_x>0$, $v_y=-v_x$ and $v_z=0$. In the above-shown panels, from left to right, velocity with respect to speed of light ($c=1$) grows according to the following sequence: $v_x=0.1$, $0.3$, and $0.7$, respectively. Larger velocity causes the accumulation of field lines and, hence, stronger intensity of the magnetic field on the front side \citep{kopacek2011}.}
\label{fig2}
\end{figure}

\section{A hidden population of isolated neutron stars in NSC}
Despite numerous observational studies of the NSC, unavoidable end-products of stellar evolution -- white dwarfs, neutron star, and stellar black holes -- still escape detection. The Galactic centre population of neutron stars is thought to be considerably numerous: order of $10^3$--$10^4$ members can be expected in the innermost parsec. This estimate is based upon the assumption of a power-law initial mass function (IMF; either standard Salpeter or the top-heavy profile) and the dynamical mass segregation over $\simeq10\,$Gyr age of the Galactic Bulge \citep{1993ApJ...408..496M}. Based on multi-wavelength statistics, \citet{2012ApJ...753..108W} reach similar order-of-magnitude estimates for the neutron star population. Using the total X-ray luminosity in the innermost parsec, \citet{2007MNRAS.377..897D} placed an upper limit on the number of compact remnants about 40 thousand members. This motivates us to explore the interaction modes of neutron stars that pass through the Galactic centre gaseous environment \citep{2015AcPol..55..203Z}. Indeed, the presence of denser and cooler gas has been reported in the Mini-spiral streamers \citep{2012A&A...538A.127K,2015AcPol..55..203Z,2017IAUS..322..129M}. 

Our region of interest is confined within the central SMBH sphere of gravitational influence $r_{\rm s}$, eq.~(\ref{eq_sphere_influence}). Passages across regions of varying density and temperature can lead to bow-shock formation and transient behaviour \citep[switching between different regimes, pulsar vs.\ accretor;][]{lipunov1992}. Other parameters influencing the interaction modes are the neutron star rotational period and its linear (quasi-Keplerian) velocity in the orbit around SMBH.

Let us consider a power-law distribution of stars inside the sphere of influence, $n=n_0 (r/r_0)^{-\gamma}$, where $\gamma$ is a power-law slope ($0<\gamma< 3$) and $n_0=n(r_0)$ number density at a reference radius $r=r_0$. This gives the radius $r_{1,{\rm NS}}$ at which the cusp of neutron star likely ends, i.e., where the estimated number of stars inside $r_{1,{\rm NS}}$ drops to unity \citep{Hopman2006ApJ...645.1152H}:
\begin{equation}
  r_{1,{\rm NS}}=(C_{\rm NS} N_{\rm s})^{-1/(3-\gamma_{\rm NS})} r_{\rm s}\,,
  \label{eq_ns_innerradius}
\end{equation}
where $N_{\rm s}$ is a number of main-sequence stars inside the sphere of influence $r_{\rm s}$, $C_{\rm NS}$ is a fraction of neutron stars, and $\gamma_{\rm NS}$ is a power-law slope of the neutron-star population. Considering the results of detailed dynamical calculations \citep{Hopman2006ApJ...645L.133H}, we take $N_{\rm s}=3.4\times 10^6$, $C_{\rm NS}=0.01$, $\gamma_{\rm NS}=1.5$, and $r_{\rm s}=1.7\,{\rm pc}$ (see Eq.~\ref{eq_sphere_influence}), which gives $r_{1,{\rm NS}}\approx 1.6 \times 10^{-3}\,{\rm pc}=8572\,r_{\rm g}$. Hence, neutron stars may be missing inside the sphere $r\leq r_{1,{\rm NS}}$. The typical dynamical time-scale, on which we can statistically expect the crossing of a neutron star through the innermost region is
\begin{equation}
   t_{\rm dyn}=\frac{2r_{1,{\rm NS}}}{v_{1,{\rm NS}}}\approx 0.95 \left(\frac{r_{1,{\rm NS}}}{1.6 \times 10^{-3}\,{\rm pc}}\right)^{3/2} \left(\frac{M_{\bullet}}{4\times 10^6\,M_{\odot}}\right)^{-1/2}\,{\rm yr}\,,
   \label{eq_dynamical_time}
\end{equation}  
where $v_{1,{\rm NS}}$ is the circular velocity at the distance of $r_{1,{\rm NS}}$. Under the assumption of the dipole field, the contribution of a neutron star at the distance $r_{1,{\rm NS}}\approx 1.6 \times 10^{-3}\,{\rm pc}$ to the magnetic field intensity at the position of Sgr~A* is negligible. Even at the distance of the order of a gravitational radius, the intensity at the position of Sgr~A* from a strongly magnetized neutron star -- magnetar -- with the magnetic moment $\mu_{\star}\approx 10^{32}\,{\rm G\,cm^{3}}$ is
\begin{equation}
  B=1\,\left(\frac{\mu_{\star}}{10^{32}\,{\rm G\,cm^{3}}}\right)\left(\frac{r}{1\,r_{\rm g}}\right)^{-3}\,{\rm mG}\,.
\end{equation}
Assuming the dipole-type structure of a magnetar magnetic field, we obtain for the magnetic intensity at $r=r_{\rm g}$,
\begin{equation}
  B_{\rm dip|r=r_{\rm g}}=1\,\left(\frac{B_{\star}}{10^{14}\,{\rm G}}\right)\left(\frac{r_{\star}}{10\,{\rm km}}\right)^{3}\left(\frac{M}{4\times10^6M_{\odot}}\right)\,{\rm mG}\,,
\end{equation}
 which is about four orders of magnitude smaller than the magnetic field intensity associated with the flare events around Sgr~A*  \citep{Eckart2012}; nonetheless, it can be comparable and even stronger than the large-scale ordered magnetic field associated with the Galactic centre non-thermal filaments \citep{2014IAUS..303..369M}. Moreover, in the course of the subsequent approach of the NS towards the SMBH, the magnetic intensity imposed onto the black hole gradually grows and it exceeds $1\;$G as the star sinks to distance of about $10^6\;$km, i.e., roughly equal to the horizon radius itself.

\section{Magnetic tubes near a black hole in motion}
We have adopted \citet{1963PhRvL..11..237K} rotating black-hole solution to Einstein field equations (\ref{eq1}) together with \citet{1985MNRAS.212..899B} solution of Maxwell equations for a weak, asymptotically uniform magnetic field, eq.\ (\ref{eq2}), i.e.\ linearized with respect to $\mathcal E$ and $\mathcal B$. We performed a translatory boost to take linear motion into account. This can be achieved by a straightforward (albeit tedious) calculation of the Lorentz transformation of the field components, which rotates the asymptotical direction of the field lines. Unlike the case of a non-drifting black hole, an electric component is induced.

The resulting magnetic field is plotted in figure \ref{fig2} for three cases of boost velocity. Let us note that the shape of lines of force depends on the choice of reference frame; we employ zero-angular-momentum observers (ZAMO) as preferred fiducial observers for which the lines of force have been constructed in a stereometric projection. We notice that the translational motion of the black hole enriches the resulting structure very significantly. We can observe formation of a boundary zone, where narrow layers surround the event horizon. Such layers are typical sites of magnetic reconnection and the subsequent acceleration of electrically charged particles.

So far there have been just very few reports of neutron star candidates near Galactic centre, which has led to a persisting ``missing pulsar problem" \citep{2014ApJ...783L...7D}. However, among confirmed pulsar detections and other evidence for magnetized neutron stars we can mention two highly dispersed pulsars at an angular offset of $\la0.3$ deg from Sgr A* \citep{2006MNRAS.373L...6J}. Also, \citet{2009ApJ...702L.177D} reported the detection of three pulsars with a large dispersion measure, indicating the location at a similar distance from us as the Galactic centre itself.

The above-given arguments motivate a scenario where a SMBH is embedded in an organized, uniform-like filament of an externally generated magnetic field. However, because of fast linear speed of the source (given by the orbital motion of the neutron star according to our picture), we need to take the effect of the translatory boost into account. To this end we adopted a suitable formalism to treat weak magnetic fields near the black hole in arbitrary motion \citep{2009CQGra..26b5004K,2009IAUS..259..127K}. Let us note that a complementary approach has been applied by \citet{2013PhRvD..88f4059D}, who explored the magnetic field of a moving dipole near a black hole within Rindler's approximation. Both methods reveal the interaction between electromagnetic and gravitational effects that are inherent to the coupled system of Einstein-Maxwell equations.

To summarize, we have illustrated the structure of an initially organized (uniform-like) magnetic lines close to the horizon, where they can be distorted by the translatory motion of the supermassive black hole. Despite a complicated structure of the field lines, a clear degree of self-organization persists, which distinguishes the field structures discussed here from highly turbulent small-scale magnetic fields that are thought to develop within accretion flows.

\section{Conclusions}
We discussed the possibility that strongly magnetized neutron stars can impose an ordered (organized) component of their dipole-type field onto a nearby supermasive black hole. To this end we discussed a likely scenario with numerous neutron stars produced as end-points of stellar evolution within the central Nuclear Star Cluster, such as the one surrounding Sgr~A* in the Galactic centre. The intensity of the dipole-type field decreases with the distance from the neutron-star surface with cube of radius in the near zone. Hence, the resulting component of the organized field comes out rather weak, although the realistic situation gets complicated beyond the light cylinder where the field lines open up and the magnetic intensity decays with radius in slower pace. Some members of the neutron star population are expected to sink very close to the black hole or even plunge into the horizon as a result of the orbital decay caused by hydrodynamical friction with the ambient gaseous environment and the emission of gravitational waves \citep[][and references cited therein]{2000ApJ...536..663N,2001A&A...376..686K}. Because of their compactness, neutron stars are not disrupted by SMBH tidal forces and they can support the magnetic intensity at and above $1\;$G on the SMBH horizon during the final stage of the infall.

\acknowledgements
VK and JK acknowledge the Czech Science Foundation grant No.\ 17-16287S titled ``Oscillations and Coherent Features in Accretion Disks around Compact Objects and Their Observational Signatures''. OK and AA thank the project LD15061 of the Czech Ministry of Education, Youth and Sports to support the collaboration within the EU COST Action MP1304 titled ``Exploring Fundamental Physics with Compact Stars''. MZ is a member of the International Max Planck Research School for Astronomy and Astrophysics at the Universities of Bonn and Cologne. 

\bibliography{karas_caosp-2017}

\end{document}